\documentclass[twocolumn,showpacs,10pt]{revtex4}

\usepackage{graphicx}
\usepackage{dcolumn}
\usepackage{bm}

\input epsf

\begin{document}

\title{\Large Thermodynamic Black Hole with Modified Chaplygin Gas\\ as a Heat Engine}

\author{\bf Ujjal Debnath\footnote{ujjaldebnath@gmail.com}}

\affiliation{Department of Mathematics, Indian Institute of
Engineering Science and Technology, Shibpur, Howrah-711 103,
India.\\}

\date{\today}

\begin{abstract}
We assume that the negative cosmological constant as a
thermodynamical pressure and the asymptotically anti-de Sitter
(AdS) black hole thermodynamics with modified Chaplygin gas. We
have written the mass of the black hole, volume, entropy and
temperature due to the thermodynamic system. We find a new
solution of Einstein's field equations of AdS black hole with
modified Chaplygin gas as a thermodynamic system. We also examine
the weak, strong and dominant energy conditions for the source
fluid of black hole. We also show that the thermodynamic black
hole with Chaplygin gas can be considered as a heat engine and
then we calculate work done and its efficiency by this system.
\end{abstract}


\maketitle

\section{\normalsize\bf{Introduction}}

Thermodynamic properties of black hole have been studied for many
years. In recent years there is considerable interest in the
physics of asymptotically AdS black hole \cite{Gibb} due to
AdS/CFT correspondence. Hawking et al \cite{Hawking} studied the
thermodynamic properties of non-rotating uncharged
Schwarzschild-AdS black hole. After that Chamblin et al
\cite{Cham1,Cham2} investigated the first order phase transition
in the non-rotating charged Reissner-Nordstrom-AdS black hole.
When the charge and/or rotation of the AdS black hole are
included, the behaviour of the AdS black hole is qualitatively
similar to the Van der Walls fluid \cite{Cv,Niu}. The concepts of
black holes from the viewpoint of chemistry, in terms of concepts
such as Van der Waals fluids, reentrant phase transitions, and
triple points have been studied in \cite{Kubi0}. The Van dar Waals
black hole has been determined by Rajagopal et al \cite{Raja}.
Subsequently, the Van dar Waals black hole in $d$-dimensions has
been described by Delsate et al \cite{Del}. Also the polytropic
black hole has been formulated by Setare et al \cite{Set}.
Kubiznak et al \cite{Kubi} assumed that the cosmological constant
$\Lambda<0$ \cite{Gun}, which represents the thermodynamic
pressure \cite{Cald,Cre,Raja,Del},
\begin{equation}
p=-\frac{\Lambda}{8\pi}=\frac{3}{8\pi l^{2}}
\end{equation}
and first law of black hole thermodynamics \cite{Raja,Del}
\begin{equation}
\delta M=T\delta S+V\delta p+.....
\end{equation}
with the black hole thermodynamics volume \cite{Raja,Del}
\begin{equation}
V=\left(\frac{\partial M}{\partial p}\right)_{S,...}
\end{equation}
where, $M$ is the mass, $S$ is the entropy and $T$ is the
temperature of the black hole. From this, an equation of state
$p=p(V,T)$ can be written for the black hole and comparing it with
the corresponding fluid equation of state, we may construct the
temperature, volume, pressure, etc.

\section{\normalsize\bf{Chaplygin Black Hole}}

Motivated by the works for Van der Waals fluid \cite{Raja,Del} and
polytropic gas \cite{Set} in AdS black hole, we assume the
modified Chaplygin gas in AdS black hole whose equation of state
is given by \cite{Debnath}
\begin{equation}
p=A\rho-\frac{B}{\rho^{\alpha}}
\end{equation}
where $A,B,\alpha$ are constants. The modified Chaplygin gas is
one of the candidate of dark energy which drives the acceleration
of the Universe. We want to construct an asymptotically AdS black
hole with Chaplygin gas whose thermodynamics coincide with the
above equation of state. So we consider the static spherically
symmetric black hole metric \cite{Raja,Del,Set}
\begin{equation}
ds^{2}=-fdt^{2}+\frac{dr^{2}}{f}+r^{2}d\Omega_{2}^{2}
\end{equation}
where
\begin{equation}
f\equiv f(r,\rho)=\frac{r^{2}}{l^{2}}-\frac{2M}{r}-g(r,\rho)
\end{equation}
Here the unknown function $g(r,\rho)$ is to be determined. Now
assume the negative cosmological constant $\Lambda$, so the
Einstein's equations are
\begin{equation}
G_{\mu\nu}+\Lambda g_{\mu\nu}=8\pi T_{\mu\nu}.
\end{equation}
Here negative $\Lambda$ represents the vacuum pressure. The
entropy, mass, volume and temperature of the black hole are
related to the horizon radius $r_{h}$ such that
\cite{Raja,Del,Set}
\begin{equation}
S=\frac{A}{4}=\pi r_{h}^{2}~,
\end{equation}
\begin{equation}
M=\frac{4\pi}{3}~r_{h}^{3}p-\frac{1}{2}~r_{h}g(r_{h},\rho)~,
\end{equation}
\begin{equation}
V=\frac{\partial M}{\partial
p}=\frac{4\pi}{3}~r_{h}^{3}-\frac{1}{2}~r_{h}\frac{\partial
g(r_{h},\rho)}{\partial\rho}/\frac{dp}{d\rho}~,
\end{equation}
\begin{eqnarray*}
T=\frac{1}{4\pi}~\left[\frac{\partial f(r,\rho)}{\partial
r}\right]_{r=r_{h}}~~~~~~~~~~~~~~~~~~~
\end{eqnarray*}
\begin{equation}
~~~~~=2r_{h}p-\frac{g(r_{h},\rho)}{4\pi
r_{h}}-\frac{1}{4\pi}~\left[\frac{\partial g(r,\rho)}{\partial
r}\right]_{r=r_{h}}
\end{equation}
Now assume that all the thermodynamic parameters for black hole
related with the parameters for modified Chaplygin gas. So the
first law of thermodynamics yields (using integrability condition)
\cite{Set}
\begin{equation}
S=\frac{\rho+p}{T}~V
\end{equation}
Using equations (4) and (8)-(11), the equation (12) reduces to
\begin{eqnarray*}
\left[24\pi
r^{2}(A\rho-B\rho^{-\alpha})-3g(r,\rho)-3r\frac{\partial
g(r,\rho)}{\partial r}\right]
\end{eqnarray*}
\begin{eqnarray*}
-\left[(1+A)\rho-B\rho^{-\alpha}\right]\times
\end{eqnarray*}
\begin{equation}
\times\left[16\pi r^{2}-6\frac{\partial g(r,\rho)}{\partial
\rho}\left\{A+\alpha B\rho^{-1-\alpha}\right\}^{-1} \right] =0
\end{equation}
Since $g(r,\rho)$ is unknown function of $r$ and $\rho$, so
without any loss of generality, we may assume the polynomial form
of $g(r,\rho)$ as in the following form
\begin{equation}
g(r,\rho)=X(r)+Y(r)\rho+Z(r)\rho^{-\alpha}
\end{equation}
where $X(r)$, $Y(r)$ and $Z(r)$ are arbitrary functions of $r$.
Now substituting the expression of $g(r,\rho)$ in equation (13),
we obtain the following equation
\begin{eqnarray*}
F_{0}(r)+F_{1}(r)\rho+F_{2}(r)\rho^{-\alpha}+F_{3}(r)\rho^{-\alpha-1}
\end{eqnarray*}
\begin{equation}
~~~~~~~~~~~~~~~~~~~~~+F_{4}(r)\rho^{-2\alpha-1}=0
\end{equation}
where
\begin{equation}
F_{0}(r)=-3A(X(r)+rX'(r)),
\end{equation}
\begin{equation}
F_{1}(r)=8A(A-2)\pi r^{2}+3(A+2)Y(r)-3ArY'(r),
\end{equation}
\begin{eqnarray*}
F_{2}(r)=-8B\pi(A+2\alpha-A\alpha)r^{2}-3B(\alpha+2)Y(r)
\end{eqnarray*}
\begin{equation}
-3(A+2\alpha+2A\alpha)Z(r)-3B\alpha rY'(r)-3ArZ'(r),
\end{equation}
\begin{equation}
F_{3}(r)=-3B\alpha (X(r)+rX'(r)),
\end{equation}
\begin{equation}
F_{4}(r)=-8B^{2}\alpha\pi r^{2}+3B\alpha Z(r)-3B\alpha rZ'(r)
\end{equation}
From the identity equation (15), comparing the co-efficients of
powers of $\rho$ in both sides, we must get
$F_{i}(r)=0,~i=0,1,2,3,4$.  Now set $F_{0}(r)=0=F_{3}(r)$, we get
(from equations (16) and (19))
\begin{equation}
X(r)=\frac{X_{0}}{r}
\end{equation}
where $X_{0}$ is an integration constant. Next put $F_{1}(r)=0$ in
equation (17), we have
\begin{equation}
Y(r)=\frac{8\pi}{3}~Ar^{2}-Y_{0}r^{1+\frac{2}{A}}
\end{equation}
where $Y_{0}$ is an integration constant. Again put $F_{4}(r)=0$
in equation (20), we must get
\begin{equation}
Z(r)=-\frac{8\pi}{3}~Br^{2}-Z_{0}r
\end{equation}
where $Z_{0}$ is another integration constant. Lastly, we set
$F_{3}(r)=0$, we obtain (from equation (18)) the relation between
two parameters as
\begin{equation}
A=-\frac{\alpha}{1+\alpha}
\end{equation}
For non-trivial solutions of $X,Y,Z$, we must have $\alpha\ne
0,A\ne 0,B\ne 0$. Putting the solutions of $X(r),~Y(r)$ and $Z(r)$
in equation (14), we get the expression of $g(r,\rho)$ as in the
following form:
\begin{eqnarray*}
g(r,\rho)=\frac{X_{0}}{r}+\left(\frac{8\pi}{3}~Ar^{2}-Y_{0}r^{1+\frac{2}{A}}\right)\rho
\end{eqnarray*}
\begin{equation}
~~~~~~~~~~+\left(-\frac{8\pi}{3}~Br^{2}-Z_{0}r\right)\rho^{-\alpha}
\end{equation}
Finally, putting the expression of $g(r,\rho)$ in equation (6), we
obtain the solution of the function $f(r,\rho)$:
\begin{equation}
f(r,\rho)=-\frac{2M+X_{0}}{r}+Y_{0}r^{1+\frac{2}{A}}\rho+Z_{0}r\rho^{-\alpha}
\end{equation}
This is a new form of black hole solution which may be called {\it
Chaplygin black hole} (after the names of Van der Waals black hole
\cite{Raja} and polytropic black hole \cite{Set}). Since the
thermodynamic pressure $p=\frac{3}{8\pi l^{2}}$ depends on $l^{2}$
and compare this pressure with the fluid pressure (eq.(4)), we may
obtain the expression of density $\rho$, which also depends on
$l^{2}$ and $\rho$ can be written in terms of pressure $p$
explicitly for some suitable values of $\alpha$. So from equation
(27), we have
\begin{equation}
f=-\frac{2M+X_{0}}{r}+Y_{0}r^{1+\frac{2}{A}}\rho+Z_{0}r\frac{A\rho-p}{B}
\end{equation}
where $\rho$ can be calculated from
$A\rho^{\alpha+1}-p\rho^{\alpha}-B=0$ with $p=\frac{3}{8\pi
l^{2}}$. So from equation (27), we may say that the black hole
solution depends of $r$ and thermodynamic pressure $p$ (which is
obviously a constant). In particular, if we choose $\alpha=-2/3$,
$X_{0}=Z_{0}=0$ and $Y_{0}=\frac{8\pi p}{\rho}$, then from
equations (25) and (27), we obtain
\begin{equation}
f=\frac{r^{2}}{l^{2}}-\frac{2M}{r}
\end{equation}
which is a black hole solution with asymptotically AdS
spacetime.\\

Now we examine the weak, strong and dominant energy conditions for
the source fluid. The energy momentum tensor for the anisotropic
source fluid is given by \cite{Raja,Del,Set}
\begin{equation}
T^{\mu\nu}=\varrho
e_{0}^{\mu}e_{0}^{\nu}+\sum_{i=1}^{3}p_{i}e_{i}^{\mu}e_{i}^{\nu}
\end{equation}
where $\varrho$ is the energy density, $p_{i}$ ($i=1,2,3$) are the
pressures for the source fluid and $e_{i}^{\mu}$ are the
components of the vielbein. Now for the black hole metric (5),
using the Einstein's equation (7), we obtain the field equations
\cite{Raja,Del,Set} (assume that the gravitational constant $G=1$)
\begin{eqnarray*}
\varrho=-p_{1}=\frac{1-f-rf'}{8\pi r^{2}}+p
~~~~~~~~~~~~~~~~~~~~~~~~~~~~~~~~~
\end{eqnarray*}
\begin{equation}
=\frac{1}{8\pi
r^{2}}\left[1+\frac{2}{\alpha}~Y_{0}r^{1+\frac{2}{A}}\rho-2Z_{0}r\rho^{-\alpha}\right]+\frac{3}{8\pi
l^{2}}
\end{equation}
and
\begin{eqnarray*}
p_{2}=p_{3}=\frac{rf''+2f'}{16\pi
r}-p~~~~~~~~~~~~~~~~~~~~~~~~~~~~~~~~~~~
\end{eqnarray*}
\begin{equation}
=\frac{1}{8\pi
r^{2}}\left[\frac{1}{\alpha}\left(1+\frac{2}{\alpha}\right)Y_{0}
r^{1+\frac{2}{A}}\rho+Z_{0}r\rho^{-\alpha}\right]-\frac{3}{8\pi
l^{2}}
\end{equation}
Now it is easy to check that the weak energy condition:
$\varrho\ge 0$, $\varrho+p_{i}\ge 0$ $(i=1,2,3)$ may be satisfied
for $Y_{0}\ge 0$, $Z_{0}\le 0$ and $\alpha>0$. The strong energy
condition: $\varrho+\sum_{i}p_{i}\ge 0$, $\varrho+p_{i}\ge 0$ may
be satisfied for $Y_{0}\ge 0$, $Z_{0}= 0$ and $\alpha\ge -2/3$.
The dominant energy condition: $\varrho\ge |p_{i}|$, $(i=1,2,3)$
may be satisfied for $Y_{0}\ge 0$, $Z_{0}\le 0$ and $0<\alpha\le
2$. So all the energy conditions will be satisfied at a time if
$Y_{0}\ge 0$, $Z_{0}= 0$ and $\alpha>0 $. In other cases, the
above energy conditions may be violated. If we assume the above
conditions of the parameters, we may checked that the energy
conditions are satisfied on the horizon. For Van der Waals black
hole \cite{Raja, Del}, some of the energy conditions are violated
but for polytropic black hole \cite{Set}, all the energy
conditions are satisfied. In our Chaplygin black hole, some of the
energy conditions are satisfied for some restrictions of the
parameters involved.

\section{\normalsize\bf{Classical Heat Engine}}

In thermodynamics and engineering, a {\it heat engine} is a system
that converts heat or thermal energy and chemical energy to
mechanical energy, which can then be used to do mechanical work.
That means a heat engine is a physical system that takes heat from
hot reservoir and part of it converts into the works while the
remaining is transferred to cold reservoir. In 2014, Johnson
\cite{John} has introduce the holographic heat engine for black
hole, where the cosmological constant was considered a
thermodynamic variable. Based on the holographic heat engine for
black hole proposal, Johnson \cite{Joh,Joh1,Joh2} has studied the
Gauss-Bonnet black holes, Born-Infeld AdS black holes and
holographic heat engines beyond large $N$ and the exact efficiency
formula. Heat engines for dilatonic Born-Infeld black holes have
been analyzed in \cite{Bha}. Zhang et al \cite{Zhan} have studied
the $f(R)$ black holes as heat engines. The thermodynamic
efficiency in charge rotating and dyonic black holes has been
studied in \cite{Sade}. Till now, several authors have studied the
heat engine mechanism for various types of black holes
\cite{Chakra,Mo,Hen,Liu,Jo,Hu1,Mo1,Hendi,Wei1,Avik,Fang,Zhang,Rosso,
Moo,Pana,Graca,J,Hhu,Santo,Fern,Z,Gha}. Recently Setare et al
\cite{Set1} have discussed polytropic black hole as a heat engine.
Motivated by their work, here we'll
study the classical heat engine for our Chaplygin black hole.\\

The horizon radius $r_{h}$ can be found from the equation
$Y_{0}r_{h}^{2+\frac{2}{A}}\rho+Z_{0}r_{h}^{2}\rho^{-\alpha}-X_{0}-2M=0$,
which depends on $X_{0},Y_{0},Z_{0},\alpha,\rho$. From equations
(8) and (10), we obtain the volume

\begin{equation}
V=\frac{Y_{0}\left(\frac{S}{\pi}\right)^{1+\frac{1}{A}}-\frac{\alpha
Z_{0}}{\pi}\rho^{-\alpha-1}S }{2(A+B\alpha \rho^{-\alpha-1})}
\end{equation}
Also from equation (11), we get the temperature
\begin{equation}
T=\left(1+\frac{1}{A}\right)\frac{2M+X_{0}}{2S}-\frac{Z_{0}(A\rho-p)}{2\pi
AB}
\end{equation}
which can be written as
\begin{equation}
S=\frac{\pi(1+A)(2M+X_{0})}{2\pi AT+Z_{0}\rho^{-\alpha-1}}
\end{equation}
So the relation between $V$ and $T$ is obtained as in the form
\begin{equation}
V=\frac{Y_{0}\left(\frac{(1+A)(2M+X_{0})}{2\pi
AT-Z_{0}\rho^{-\alpha-1}}\right)^{1+\frac{1}{A}}- \frac{\alpha
Z_{0}(1+A)(2M+X_{0})\rho^{-\alpha-1}}{2\pi
AT-Z_{0}\rho^{-\alpha-1}}}{2(A+B\alpha \rho^{-\alpha-1})}
\end{equation}
To describe the thermodynamic behavior of the Chaplygin gas in
presence of variable pressure (i.e., variable cosmological
constant), one can identify mass $M$ from being the energy $U$ to
being the {\it enthalpy} \cite{Kastor}, i.e., the enthalpy
function is defined by $H = M = U + pV$. From the first law of
thermodynamics, we get
\begin{equation}
dH=dM=TdS+Vdp
\end{equation}
By integration the above equation, the enthalpy function can be
written in the form:
\begin{equation}
H=-\frac{1}{2}X_{0}+\frac{Y_{0}\rho}{2}\left(\frac{S}{\pi}
\right)^{1+\frac{1}{A}}+\frac{Z_{0}S}{2\pi\rho^{\alpha}}
\end{equation}
The Gibb's free energy is given by \cite{Graca}
\begin{eqnarray}
G=H-TS =-\frac{1}{2}X_{0}-\frac{Y_{0}\rho}{2A}\left(\frac{S}{\pi}
\right)^{1+\frac{1}{A}}
\end{eqnarray}
Also the free energy is given by \cite{Graca}
\begin{eqnarray}
F=G-pV=-\frac{1}{2}X_{0}-\frac{Y_{0}\rho}{2A}\left(\frac{S}{\pi}
\right)^{1+\frac{1}{A}} \nonumber\\
-\frac{pY_{0}\left(\frac{S}{\pi}\right)^{1+\frac{1}{A}}-\frac{\alpha
Z_{0}}{\pi}\rho^{-\alpha-1}pS }{2(A+B\alpha \rho^{-\alpha-1})}
\end{eqnarray}

David Kubiznak and Robert B. Mann \cite{Kubiz} have showed the
critical behaviour of charged AdS black holes. Following this, we
will study the critical behavior of the Chaplygin black hole.
Critical point is a point of inflection which can be found from
the following conditions:
\begin{equation}
\left(\frac{\partial p}{\partial r_{h}}
\right)_{cr}=0,~~~~\left(\frac{\partial^{2} p}{\partial r_{h}^{2}}
\right)_{cr}=0
\end{equation}
At the critical point $r_{cr}$, the critical pressure $p_{cr}$ and
critical temperature $T_{cr}$ will be
\begin{equation}
p_{cr}=A\left(\frac{Y_{0}}{\alpha
Z_{0}}\right)^{-\frac{1}{\alpha+1}}
r_{cr}^{\frac{2}{\alpha}}-B\left(\frac{Y_{0}}{\alpha
Z_{0}}\right)^{\frac{\alpha}{\alpha+1}} r_{cr}^{-2}
\end{equation}
and
\begin{equation}
T_{cr}=\left[\frac{2M+X_{0}}{2\pi \alpha} +\frac{Z_{0}}{2\pi
A}\left(\frac{Y_{0}}{\alpha
Z_{0}}\right)^{\frac{\alpha}{\alpha+1}}\right] r_{cr}^{-2}
\end{equation}
with the condition
\begin{equation}
2M+X_{0}=\frac{Y_{0}(\alpha+1)}{\alpha}\left(\frac{Y_{0}}{\alpha
Z_{0}}\right)^{-\frac{1}{\alpha+1}}
\end{equation}

Now assume, $T_{H}$ and $T_{C}$ are the temperatures of the hot
and cold reservoirs respectively and they consist of two
isothermal processes with two adiabatic processes. The heat engine
flow is shown figure 1 \cite{John}. So the heat flow for the upper
isotherm process from 1 to 2 is given by \cite{John}
\begin{equation}
Q_{H}=T_{H}\triangle S_{1\rightarrow 2}=T_{H}(S_{2}-S_{1})
\end{equation}
and the exhausted heat from the lower isothermal process is given
by \cite{John}
\begin{equation}
Q_{C}=T_{C}\triangle S_{3\rightarrow 4}=T_{C}(S_{3}-S_{4})
\end{equation}
Here $S_{i}$'s are related to $V_{i}$'s satisfying
\begin{equation}
V_{i}=\frac{Y_{0}\left(\frac{S_{i}}{\pi}\right)^{1+\frac{1}{A}}-\frac{\alpha
Z_{0}}{\pi}\rho_{i}^{-\alpha-1}S_{i} }{2(A+B\alpha
\rho_{i}^{-\alpha-1})}~,~i=1,2,3,4.
\end{equation}
where $\rho_{i}$ can be calculated from the relation
$p_{i}=A\rho_{i}-B\rho_{i}^{-\alpha}$,~$i=1,2,3,4$. The $p$-$V$
diagram \cite{John} shows the Carnot heat engine which forms a
closed path in figure 2.

The work done by the heat engine is
\begin{equation}
W=Q_{H}-Q_{C}
\end{equation}
The efficiency of a heat engine relates how much useful work is
output for a given amount of heat energy input and it is defined
by
\begin{equation}
\eta=\frac{W}{Q_{H}}=1-\frac{Q_{C}}{Q_{H}}
\end{equation}
We know that the Carnot cycle has the maximum efficiency. Also we
mention that the Stirling cycle consists of two isothermal
processes plus two isochores processes. So the maximally efficient
Carnot engine is also a Stirling engine. For Carnot cycle,
$V_{1}=V_{4}$ and $V_{2}=V_{3}$, so we have the maximum efficiency
as
\begin{equation}
\eta_{max}=1-\frac{T_{C}}{T_{H}}
\end{equation}
which is the maximum one of all the possible cycles between the
given higher temperature $T_{H}$ and lower one $T_{C}$. The
specific heat of the thermodynamical system is
\begin{eqnarray}
C=T\left(\frac{\partial S}{\partial
T}\right)=-\frac{2AS^{2}T}{(1+A)(2M+X_{0})} \times \nonumber \\
 \left[1-\frac{Z_{0}\alpha \rho^{-\alpha-1}}{2\pi A(A+B\alpha
\rho^{-\alpha-1}) }~ \frac{\partial p}{\partial T}\right]
\end{eqnarray}
If volume $V$ is constant (i.e., $S$ is constant), the we can
obtain
\begin{equation}
\left(\frac{\partial p}{\partial T}\right)_{V}=\frac{2\pi
A(A+B\alpha \rho^{-\alpha-1})}{Z_{0}\alpha \rho^{-\alpha-1}}
\end{equation}
and hence consequently for constant volume, the specific heat
$C_{V}=0$. For constant pressure, $\left(\frac{\partial
p}{\partial T}\right)_{p}=0$, so we may obtain the specific heat
for constant pressure as
\begin{equation}
C_{p}=-\frac{2AS^{2}T}{(1+A)(2M+X_{0})}
\end{equation}
which is not equal to zero. So we have a new engine, described in
figure 3, which involves two isobars and two isochores/adiabats
\cite{John}. The heat flows show along the top and bottom. The
work done along the isobars is given by
\begin{eqnarray}
W=\triangle p_{4\rightarrow 1}~\triangle V_{1\rightarrow
2}=(p_{1}-p_{4})(V_{2}-V_{1}) ~~~~~~~~~~~~~~~~~~~~~~~~~ \nonumber\\
=(p_{1}-p_{4})~\left[\frac{Y_{0}}{2}\left\{\frac{\left(\frac{S_{2}}{\pi}\right)^{1+\frac{1}{A}}}{(A+B\alpha
\rho_{2}^{-\alpha-1})}
-\frac{\left(\frac{S_{1}}{\pi}\right)^{1+\frac{1}{A}}}{(A+B\alpha
\rho_{1}^{-\alpha-1})} \right\} \right.  \nonumber\\
\left. -\frac{\alpha
Z_{0}}{2\pi}\left\{\frac{\rho_{2}^{-\alpha-1}S_{2}}{(A+B\alpha
\rho_{2}^{-\alpha-1})}-\frac{\rho_{1}^{-\alpha-1}S_{1}}{(A+B\alpha
\rho_{1}^{-\alpha-1})}\right\} \right]~~~~~~~~~~~~
\end{eqnarray}
The net inflow of heat in upper isobar is given by
\begin{equation}
Q_{H}=\int_{T_{1}}^{T_{2}} C_{p}(p_{1},T)dT
\end{equation}
which can be expressed as
\begin{eqnarray}
Q_{H}= \frac{(A+1)(2M+X_{0})}{2A}~~~~~~~~~~~~~~~~~~~~~~~~~~ \nonumber\\
\times
 \left[\frac{4\pi AZ_{0}\rho^{-\alpha}(T_{2}-T_{1}) }{(2\pi
AT_{1}+Z_{0}\rho_{1}^{-\alpha})(2\pi
AT_{2}+Z_{0}\rho_{1}^{-\alpha})}\right. \nonumber \\
\left. +Log\left(\frac{ 2\pi AT_{1}+Z_{0}\rho_{1}^{-\alpha}}{2\pi
AT_{2}+Z_{0}\rho_{1}^{-\alpha}} \right)\right]~~~~~~~~~~~~~~
\end{eqnarray}
or in the other form:
\begin{equation}
Q_{H}=\frac{(1+A)(2M+X_{0})}{2A}~Log\frac{S_{2}}{S_{1}}+\frac{Z_{0}\rho_{1}^{-\alpha}}{2\pi
A}~(S_{1}-S_{2})
\end{equation}
Finally we can demonstrate the performance of the heat engine by a
thermal efficiency $\eta$ and found in the following  form:
\begin{eqnarray*}
\eta=\frac{W}{Q_{H}}=\left(1-\frac{p_{4}}{p_{1}}\right)\times
\hspace{3in}
\nonumber\\
\left[\frac{Y_{0}p_{1}}{2}\left\{\frac{\left(\frac{S_{2}}{\pi}\right)^{1+\frac{1}{A}}}{(A+B\alpha
\rho_{2}^{-\alpha-1})}
-\frac{\left(\frac{S_{1}}{\pi}\right)^{1+\frac{1}{A}}}{(A+B\alpha
\rho_{1}^{-\alpha-1})} \right\} \right. \hspace{1.5in} \nonumber\\
\left. -\frac{\alpha
Z_{0}p_{1}}{2\pi}\left\{\frac{\rho_{2}^{-\alpha-1}S_{2}}{(A+B\alpha
\rho_{2}^{-\alpha-1})}-\frac{\rho_{1}^{-\alpha-1}S_{1}}{(A+B\alpha
\rho_{1}^{-\alpha-1})}\right\} \right]\times \hspace{1.5in}
\nonumber
\end{eqnarray*}
\begin{equation}
\left[\frac{(1+A)(2M+X_{0})}{2A}~Log\frac{S_{2}}{S_{1}}+\frac{Z_{0}\rho_{1}^{-\alpha}}{2\pi
A}~(S_{1}-S_{2})\right]^{-1}
\end{equation}
which crucially depends on the modified Chaplygin gas parameters
$\alpha,~A$ and $B$.

\section{\normalsize\bf{Discussions}}

We have assumed the negative cosmological constant as a
thermodynamical pressure and the asymptotically anti-de Sitter
(AdS) black hole thermodynamic parameters which are identical with
the modified Chaplygin gas, which obeys the integrability
condition of the thermodynamical system. We have written the mass
of the black hole, volume, entropy and temperature due to the
thermodynamic system. We found the solutions of Einstein's field
equations of AdS black hole for modified Chaplygin gas. The new
form of solution for black hole may be called Chaplygin black hole
(after the names of Van der Waals black hole \cite{Raja} and
polytropic black hole \cite{Set}). For $A=0$, the above Chaplygin
black hole solution may be reduced to the polytropic black hole
solution for negative $\Lambda$. If we set $A=2$,
$Y_{0}=\frac{1}{l^{2}\rho}$ and $Z_{0}=0$, the Chaplyin black hole
may be reduced to the asymptotically AdS black hole for negative
$\Lambda$. Also if we set $A=-2$, $Y_{0}=\frac{1}{\rho}$ and
$X_{0}=Z_{0}=0$, the Chaplyin black hole may be reduced to the
Schwarszchild black hole. We have also examined the weak, strong
and dominant energy conditions for the source fluid of the
Chaplygin black hole. For $Y_{0}\ge 0$, $Z_{0}\le 0$ and
$\alpha>0$, the weak energy condition is satisfied, for $Y_{0}\ge
0$, $Z_{0}\le 0$ and $\alpha\le -2/3$, the dominant energy
condition is satisfied and for $Y_{0}\ge 0$, $Z_{0}= 0$ and
$0<\alpha\le 2$, the strong energy condition is satisfied but for
other cases, the all energy conditions are violated.\\

We have described the classical heat engine for Chaplygin black
hole. Using the horizon radius $r_{h}$, we have found the
relations between volume $V$, temperature $T$, entropy $S$ and
pressure $p$ (or density $\rho$). Using the first law of
thermodynamics, we have found the enthalpy function in terms of
the entropy $S$. The Gibb's free energy and free energy have been
evaluated. The critical pressure and critical temperature have
been found at the critical point of the system. We have found the
heat flows from upper and lower isotherms process. Also we have
calculated the work done by the heat engine and its efficiency. We
have found the maximum efficiency by the Carnot cycle. For static
black holes, Johnson \cite{John} has investigated that the Carnot
and Stirling cycles are coincided. For constant volume, we found
that specific heat $C_{V}=0$. On the other hand, for constant
pressure, we have found that specific heat $C_{p}\ne 0$. So we
have considered another cycle which consists of two isobars and
two isochores. We have calculated the net inflow of heat in upper
isobar and efficiency
of the heat engine for this cycle.\\

\begin{figure}[!h]
     \includegraphics[width=0.4\linewidth]{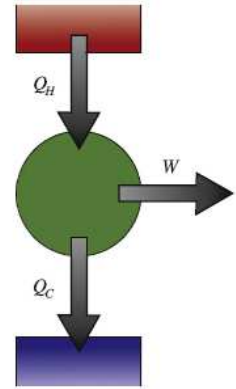}
    \caption{The figure represents the heat engine
    flows \cite{John}.}
\end{figure}

\begin{figure}[!h]
     \includegraphics[width=0.4\linewidth]{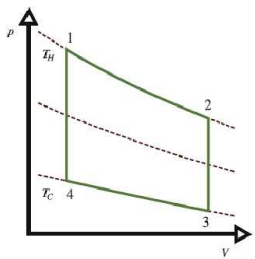}
    \caption{The figure represents the Carnot Engine \cite{John}.}
\end{figure}

\begin{figure}[!h]
     \includegraphics[width=0.4\linewidth]{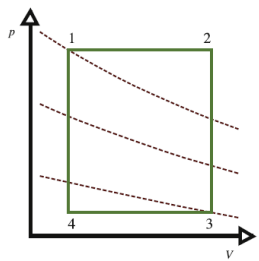}
    \caption{The figure represents other Engine \cite{John}.}
\end{figure}

\end{document}